\documentclass[extra,onecolumn]{gji}
\usepackage[fleqn]{amsmath}
\usepackage{amsfonts,amssymb}
\usepackage{timet,epsfig,wrapfig,psfrag}
\usepackage{natbib}
\usepackage{symbol}
\usepackage{color}
\usepackage[normalem]{ulem}
\usepackage{booktabs}
\usepackage{multirow} 
\usepackage{pdfcolmk}

\newcommand{\w}{\mathbf}

\begin{document}

\bibliographystyle{chicago}


\title[Sensitivity kernels for transmission fiber optics]
{Sensitivity kernels for transmission fiber optics}

\author[Andreas Fichtner et al.]{
\parbox{\linewidth}{Andreas Fichtner$^1$, Adonis Bogris$^2$, Daniel Bowden$^1$, Konstantinos Lentas$^3$, Nicos Melis$^3$, Thomas Nikas$^4$, Christos Simos$^5$, Iraklis Simos$^2$ and Krystyna Smolinski$^1$} \\
$^1$ Department of Earth Sciences, ETH Zurich, Switzerland\\
$^2$ University of West Attica, Athens, Greece\\
$^3$ National Observatory of Athens, Institute of Geodynamics, Greece\\
$^4$ Department of Informatics and Telecommunications, National and Kapodistrian University of Athens, Athens, Greece\\
$^5$ University of Thessaly, Department of Physics, Electronics and Photonics Laboratory, Lamia, Greece}

\maketitle


\begin{summary}
Fiber-optic sensing technologies based on transmission offer an alternative to scattering-based Distributed Acoustic Sensing (DAS). Being able to interrogate fibers that are thousands of kilometers long, opens opportunities for seismological studies of remote regions, including ocean basins. However, by averaging deformation along the fiber, transmission systems only produce integrated and not distributed measurements. Here we develop a formalism to calculate sensitivity kernels with respect to (Earth) structure, using optical phase delay measurements. With this, we demonstrate that transmission-based sensing can effectively provide distributed measurements when the phase delay time series is dissected into different windows. The extent to which a potentially useful sensitivity coverage can be achieved, depends on the fiber geometry, and specifically on its local curvature. This work establishes a theoretical foundation for both tomographic inversions and experimental design, using transmission-based optical sensing.
\end{summary}


\section{Introduction}\label{S:Introduction}

Distributed Acoustic Sensing (DAS) is an emerging family of technologies that permits seismic data acquisition with high spatio-temporal resolution and large bandwidth using fiber-optic cables \citep[e.g.,][]{Hartog_2017,Lindsey_2020,Paitz_2021}. The relative ease of deploying cables in challenging terrain, and opportunities to co-use existing telecommunication infrastructure, have enabled seismological applications that would have been substantially more difficult or expensive using conventional seismic instruments \citep[e.g.][]{Lindsey_2017,Martin_2017,Walter_2020,Klaasen_2021}. While Rayleigh back-scattering allows DAS to achieve distributed measurements with an effective channel spacing in the centimeter range, it also limits the length of the fiber that can be interrogated to several tens of kilometers.\\
Alternative systems based on optical transmission instead of back-scattering, overcome this limitation by measuring deformation-induced changes of phase \citep{Marra_2018,Bogris_2021,Bogris_2022,Bowden_2022} or polarization \citep{Mecozzi_2021}. Reaching interrogation distances of hundreds or thousands of kilometers, opens new opportunities to study seismic activity and Earth structure in remote regions, including ocean basins. The main drawback of transmission-based systems lies in the averaging of deformation along the fiber. Hence, in contrast to DAS, the measurement is not distributed but integrated. As a consequence, it remains unclear if and how transmission measurements may be used, for instance, to infer Earth structure.\\
In the following sections, we develop a formalism for the calculation of sensitivity kernels with respect to Earth structure based on measurements of deformation-induced optical phase delays. Using simple conceptual examples, we demonstrate that dissecting the phase delay time series can effectively enable distributed measurements, thereby providing a theoretical foundation for tomographic inversions and optimal experimental design. 


\section{Theoretical developments}\label{S:Theory}

\begin{center}
\begin{figure}
\center{\includegraphics[width=0.75\textwidth, angle=0]{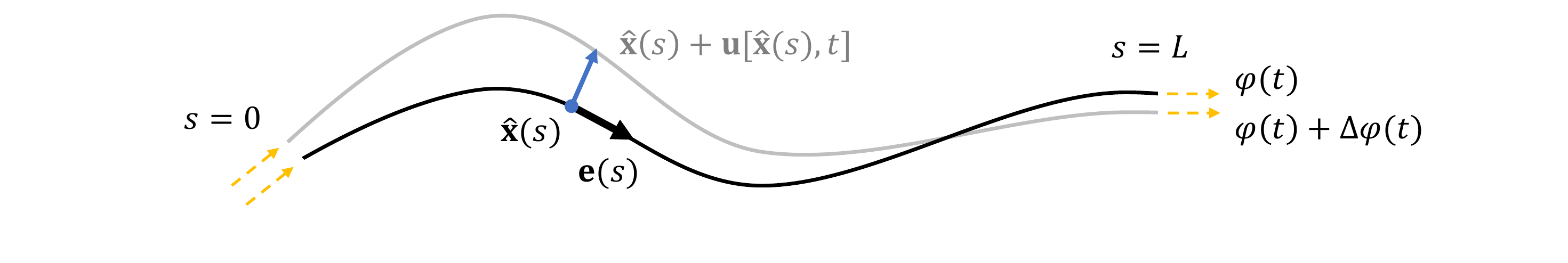}}
\caption{Schematic illustration of fiber deformation. The undeformed fiber, shown as black curve, is represented by the position vector $\hat{\mathbf{x}}(s)$, which is parametrized in terms of the arc length $s\in[0,L]$. The displacement field $\mathbf{u}[\hat{\mathbf{x}}(s),t]$ in blue moves $\hat{\mathbf{x}}(s)$ to $\hat{\mathbf{x}}(s)+\w{u}[\hat{\mathbf{x}}(s)]$. The result is the deformed fiber in gray. The local tangent vector $\mathbf{e}(s)$ is displayed as a thick black arrow. Upon deformation, an optical signal, shown in orange, acquires a phase change $\Delta\phi(t)$.}\label{F:geometry}
\end{figure}
\end{center}

We consider a fiber with position $\w{x}=\hat{\w{x}}(s)$ parametrized in terms of the arc length $s\in[0,L]$, where $L$ is the total length, as illustrated in Fig. \ref{F:geometry}. In terms of the effective refractive index $r(s)$, the vacuum speed of light $c$, and the circular frequency $\omega$ of the optical signal, its time-dependent phase change $\theta(t)=\Delta\dot{\varphi}(t)$  can be represented in two equivalent forms \citep{Fichtner_2022a,Bowden_2022},
\begin{equation}\label{E:000}
\theta(t) = \frac{\omega}{c} \int_{s=0}^L r(s) \, \dot{\varepsilon}[\hat{\w{x}}(s),t]\, ds = - \frac{\omega}{c} \int_{s=0}^L \frac{d}{ds} \left[ r(s)\,\mathbf{e}(s) \right] \cdot \dot{\w{u}}[\hat{\w{x}}(s),t]\, ds\,,
\end{equation}
where an over-dot denotes a time derivative. Eq. (\ref{E:000}) relates the optical phase change $\theta(t)$ to the deformation of the medium, expressed in terms of the displacement field $\w{u}(\w{x},t)$ or the axial component $\varepsilon=\w{e}\cdot \w{E} \cdot \w{e}$ of the strain tensor $\w{E}=(\nabla\w{u}+\nabla\w{u}^T)/2$.  The first variant of (\ref{E:000}) allows us to compare transmission measurements $\theta(t)$ to axial strain rate measurements $\dot{\varepsilon}[\hat{\w{x}}(s),t]$ from DAS. The second variant of (\ref{E:000}), which we will use in the following paragraphs, exhibits the dependence of $\theta(t)$ on fiber curvature and spatial variations of the effective refractive index.
%
%
The dynamic fields $\w{u}$ and $\varepsilon$ depend on the distribution of medium properties, such as wave speeds and density,  collected in the model vector $\w{m}(\w{x})$ and omitted in the notation to avoid clutter. To infer a plausible $\w{m}(\w{x})$ from the comparison of observed and calculated phase changes, $\theta^\text{obs}$ and $\theta$, we require a suitable measurement functional. Among numerous available options \citep[e.g.,][]{Gee_1992,Fichtner_et_al_2008,Bozdag_2011} , we choose the cross-correlation time shift within a windowed section of the phase change time series \citep{Luo_Schuster_1991}. For this, we define the windowed cross-correlation as 
\begin{equation}\label{E:001}
\mathcal{C}(\tau)=\int_{t=-\infty}^{\infty} w(t) \, \theta^\text{obs}(t)\,\theta(t+\tau)\,dt\,,
\end{equation}
where $w(t)$ denotes a window function that isolates a selected part of the time series. The time shift $\mathcal{T}$ between the windowed versions of $\theta^\text{obs}$ and $\theta$ is defined as the value of $\tau$ where $\mathcal{C}(\tau)$ reaches its global maximum. Hence, by differentiating (\ref{E:001}), we obtain
\begin{equation}\label{E:002}
0 = \dot{\mathcal{C}}(\mathcal{T}) = \int_{t=-\infty}^{\infty} w(t) \, \theta^\text{obs}(t)\,\dot{\theta}(t+\mathcal{T})\,dt =  - \int_{t=-\infty}^{\infty} \dot{\theta}_w^\text{obs}(t-\mathcal{T})\,\theta(t)\,dt\,,
\end{equation}
with the convenient definition of the windowed phase change $\theta_w^\text{obs}=w\theta^\text{obs}$. Eq. (\ref{E:002}) defines the measurement functional $\mathcal{T}$ implicitly. Variations $\delta\w{m}(\w{x})$ of medium parameters induce variations $\delta\mathcal{T}$ of the measurement. Invoking implicit function differentiation, we obtain
\begin{equation}\label{E:003}
\delta\mathcal{T} = -  \int_{t=-\infty}^{\infty} \dot{\theta}_w^\text{obs}(t-\mathcal{T})\,\delta\theta(t)\,dt \,/ \int_{t=-\infty}^{\infty} \ddot{\theta}_w^\text{obs}(t-\mathcal{T})\, \theta(t)\,dt \,.
\end{equation}
Making the common assumption that $\theta_w^\text{obs}$ is approximately a time-shifted version of $\theta_w$, i.e., $\theta_w^\text{obs}(t-\mathcal{T})\approx \theta_w(t)$ \citep[e.g.,][]{Luo_Schuster_1991,Dahlen_2000}, we can simplify Eq. (\ref{E:003}) to 
\begin{equation}\label{E:004}
\delta\mathcal{T} = ||\dot{\theta}||_w^{-2} \, \int_{t=-\infty}^{\infty}  \dot{\theta}_w(t)\,\delta\theta(t)\,dt\,,
\end{equation}
with the squared norm $||\dot{\theta}||_w^{2}=\int_{t=-\infty}^{\infty}  w(t)\,\dot{\theta}^2(t)\,dt$. To relate variations of the measurement functional $\delta\mathcal{T}$ to variations in medium parameters $\delta\w{m}(\w{x})$, we substitute the variation of (\ref{E:000}) into (\ref{E:004}), 
\begin{equation}\label{E:005}
\delta\mathcal{T} = - ||\dot{\theta}||_w^{-2} \, \int_{t=-\infty}^{\infty} \int_{s=0}^L \ddot{\theta}_w(t) \, \w{a}(s) \cdot \delta\w{u}[\hat{\w{x}}(s),t]\,ds\,,
\end{equation}
where we defined the vector $\w{a}(s)=\frac{\omega}{c} \frac{d}{ds} \left[ r(s)\,\mathbf{e}(s) \right]$ for notational convenience. Since $\delta\w{u}$ can in practice not be computed efficiently for large numbers of arbitrary perturbations $\delta\w{m}(\w{x})$, we appy the adjoint method \citep[e.g.,][]{Tromp_2005,Fichtner_2006a}. For this, we write the (seismic) wave equation symbolically in terms of the wave equation operator $\w{L}$ as $\w{L}[\w{u}(\w{m}),\w{m}]=\w{f}$, where $\w{f}$ is the source term. Multiplying the variation of the wave equation, $\delta\w{L}[\w{u}(\w{m}),\w{m}] + \w{L}[\delta\w{u}(\w{m}),\w{m}]=\w{0}$ with an arbitrary field $\w{u}^\dag$ and integrating over time and the spatial domain $\oplus \subset \mathbb{R}^3$, yields
\begin{equation}\label{E:006}
\int_{t=-\infty}^{\infty} \int_\oplus \w{u}^\dag \cdot \delta\w{L}[\w{u}(\w{m}),\w{m}] \,d^3\w{x}\,dt + \int_{t=-\infty}^{\infty} \int_\oplus \w{u}^\dag \cdot \w{L}[\delta\w{u}(\w{m}),\w{m}]\, d^3\w{x}\,dt = 0\,.
\end{equation}
Invoking the adjoint $\w{L}^\dag$ of $\w{L}$ and adding (\ref{E:005}) to (\ref{E:006}), we find
\begin{equation}\label{E:007}
\delta\mathcal{T} = \int_{t=-\infty}^{\infty} \int_\oplus \delta\w{u} \cdot \left[ \w{L}^\dag(\w{u}^\dag,\w{m}) - ||\dot{\theta}||_w^{-2} \ddot{\theta}_w(t) \, \w{a}(s) \, \delta_L(\w{x}) \right]\,d^3\w{x}\,dt + \int_{t=-\infty}^{\infty} \int_\oplus \w{u}^\dag \cdot \delta\w{L}[\w{u}(\w{m}),\w{m}]\, d^3\w{x}\,dt\,.
\end{equation}
In (\ref{E:007}), the curve $\delta$-distribution $\delta_L$ is defined as $\int_{s=0}^L f[\hat{\w{x}}(s)]\,ds = \int_\oplus f(\w{x})\,\delta_L(\w{x})\,d^3\w{x}$ for any function $f$ in $\oplus$. We can now eliminate $\delta\w{u}$ from (\ref{E:007}) by forcing the first term on the right-hand side to zero and thereby defining the adjoint field $\w{u}^\dag$ as the solution of the adjoint equation
\begin{equation}\label{E:008}
\w{L}^\dag(\w{u}^\dag,\w{m}) = ||\dot{\theta}||_w^{-2} \ddot{\theta}_w(t) \, \w{a}(s) \, \delta_L(\w{x})\,.
\end{equation}
The right-hand side of (\ref{E:008}) is the adjoint source, which $\delta_L$ localizes along the fiber. The time evolution of the adjoint source is controlled by the second derivative of the windowed computed phase changes, $\ddot{\theta}_w$, and its position-dependent orientation is given by the vector $\w{a}(s)$. What is left of Eq. (\ref{E:007}) can be conveniently written in terms of the structural sensitivity kernel $\w{K}$, i.e., the volumetric sensitivity of the time shift $\mathcal{T}$,
\begin{equation}\label{E:009}
\delta\mathcal{T} = \int_{t=-\infty}^{\infty} \int_\oplus \w{u}^\dag \cdot \delta\w{L}[\w{u}(\w{m}),\w{m}]\, d^3\w{x}\,dt\, = \int_\oplus \w{K}(\w{x})\cdot \delta\w{m}\,.
\end{equation}
Explicit expressions of $\w{K}$ for various medium parameters can be found, e.g., in \citet{Tromp_2005} or \citet{Fichtner_book}, and will not be repeated here.

\section{Conceptual examples}\label{S:Examples}

\begin{center}
\begin{figure}
\center{\includegraphics[width=0.7\textwidth, angle=0]{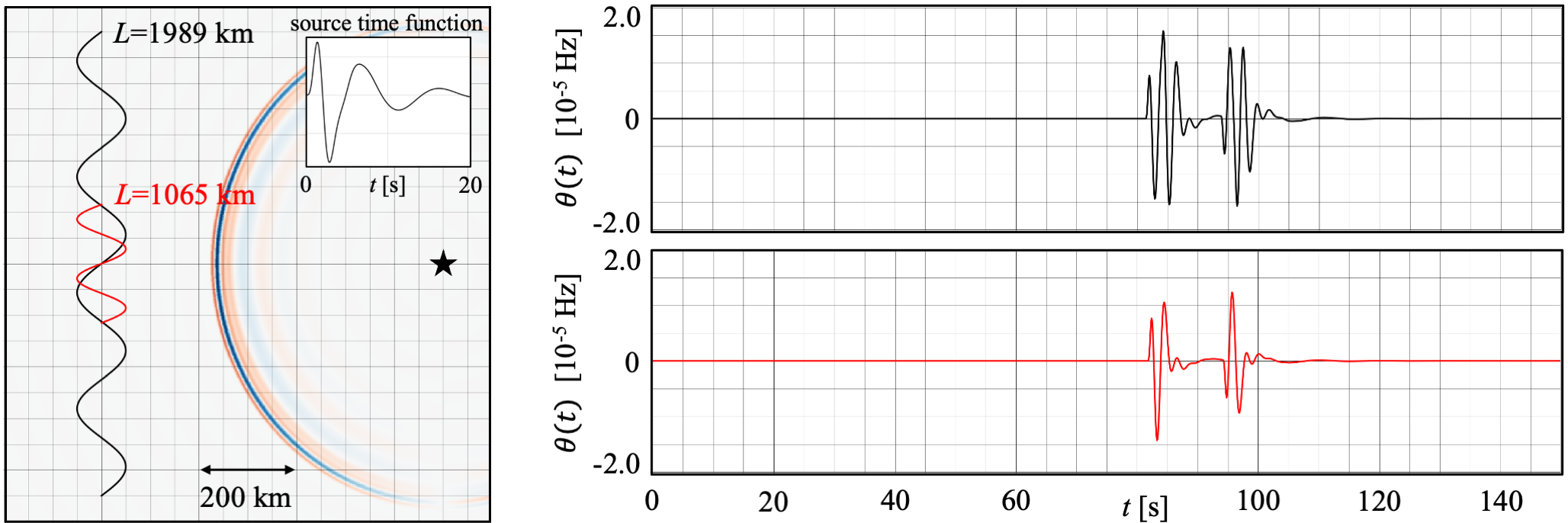}}
\caption{Forward modelling example using two different fiber geometries with lengths of $L=1989$ km (black) and $L=1065$ km (red). The wavefield radiates from an explosive source at the location of the star. A snapshot of the $x$-component is shown in the left panel, together with the source-time function; a Heaviside function filtered between $0.1$ and $0.5$ Hz. The resulting time series $\theta(t)$ are shown to the right, in the color of the corresponding fiber geometry.}\label{F:forward}
\end{figure}
\end{center}

\begin{center}
\begin{figure}
\center{\includegraphics[width=1.0\textwidth, angle=0]{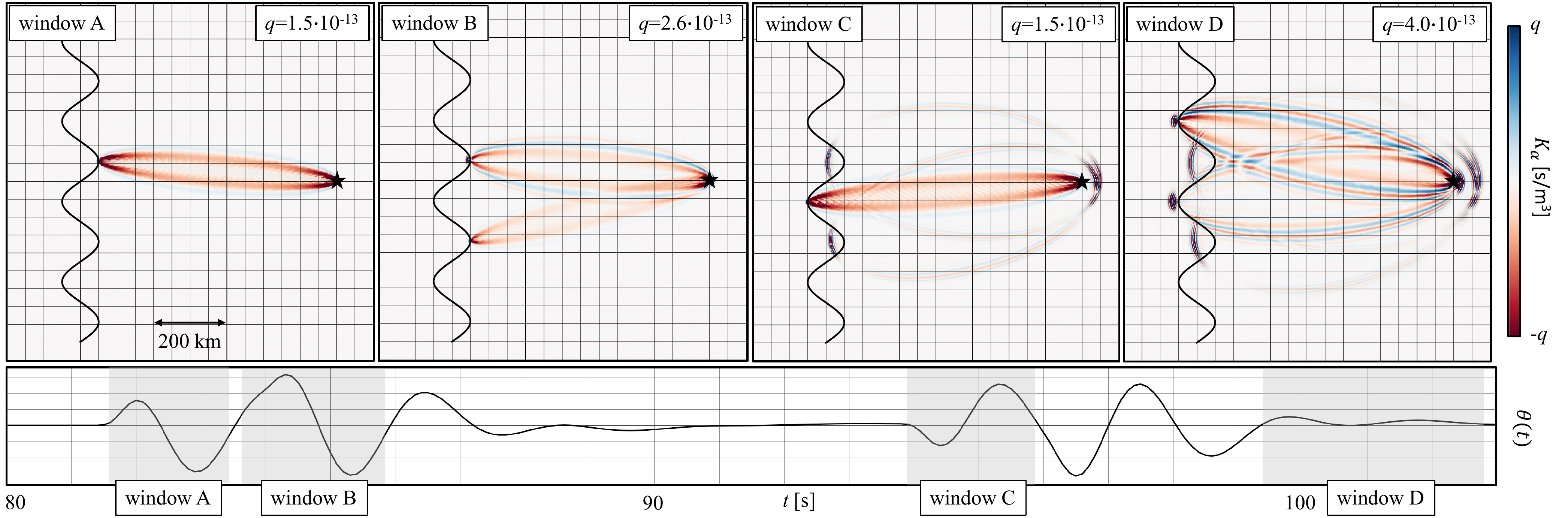}}
\caption{Sensitivity kernels $K_\alpha$ for relative P velocity perturbations $\delta\ln\alpha$. Time shift measurements $\mathcal{T}$ in different windows, shown in the lower panel, produce kernels with different spatial coverage that indicate the section of the fiber where the signal has been primarily generated.}\label{F:kernels_1}
\end{figure}
\end{center}

\begin{center}
\begin{figure}
\center{\includegraphics[width=1.0\textwidth, angle=0]{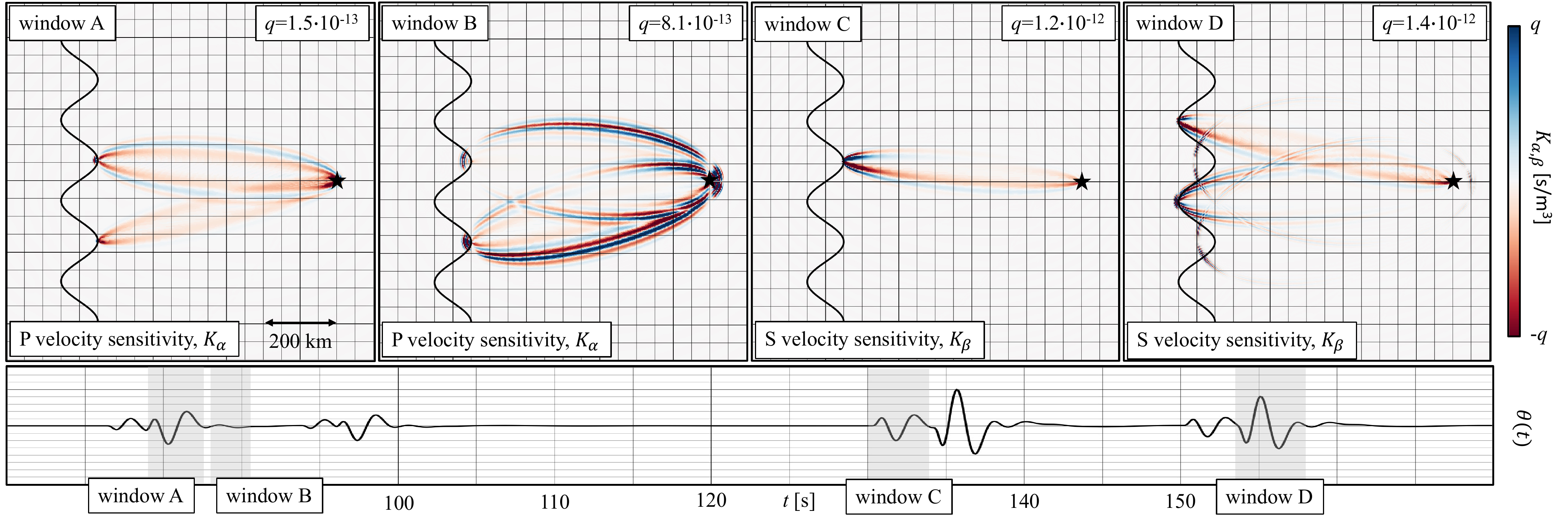}}
\caption{Sensitivity kernels $K_\alpha$ and $K_\beta$ for relative velocity perturbations $\delta\ln\alpha$ and $\delta\ln\beta$, respectively. The phase change time series $\theta(t)$ and the different measurement windows used to calculate kernels are shown in the lower panel.}\label{F:kernels_2}
\end{figure}
\end{center}

The following examples are intended to illustrate the generation of phase change signals $\theta(t)$ and their time-dependent sensitivity to medium parameters. We make the plausible assumption that the effective refractive index $r(s)$ is constant over a seismic wavelength, i.e., $\mathcal{O}(10)$ km in our examples, thereby allowing us to ignore its derivative. Not trying to mimic a specific acquisition system, we set $\omega r / c=1$ m$/$s$^2$ for simplicity. The normalization factor $||\dot{\theta}||_w^{-2}$ in the adjoint source (\ref{E:008}) ensures that traveltime kernels are unaffected by this choice, and by any other amplitude scaling, e.g., via the seismic moment of the wavefield source. With this setting, the vector $\w{a}(s)$ equals the non-normalized normal vector $\frac{d}{ds}\w{e}(s)$. To avoid any complications, the elastic medium for our calculations is unbounded, isotropic and perfectly elastic, with P velocity $\alpha=8000$ m$/$s, S velocity $\beta=5000$ m$/$s, and density $\rho=3000$ kg$/$m$^3$. Well-known analytical solutions for moment tensor and single force sources may be found, e.g., in \citet{Aki_Richards_2002}.\\
Starting with the simplest possible case, we consider an explosive source that only radiates a P wave, as shown in Fig. \ref{F:forward}. The displacement field $\w{u}(\w{x},t)$ deforms two different fibers, plotted as black and red curves, respectively. Since $\w{u}(\w{x},t)$ interacts with different parts of the fibers at different times, the resulting phase changes $\theta(t)$ contain multiple oscillations that are more complex than the seismic P wavelet. Even though the red fiber is only half as long as the black fiber, it produces phase changes of nearly the same amplitude, because $\theta(t)$ is proportional to  $\w{a}(s)=\frac{d}{ds}\w{e}(s)$, i.e., to curvature. Hence, the example in Fig. \ref{F:forward} illustrates that larger curvature may compensate shorter length, and vice versa.\\
Applying the formalism developed in section \ref{S:Theory}, allows us to compute sensitivity kernels $K_\alpha$ for relative perturbations of P velocity, $\delta\ln\alpha$, some examples of which are displayed in Fig. \ref{F:kernels_1}. Using different windows $w(t)$ for the time shift measurements $\mathcal{T}$, produces kernels with different spatial coverage. They exhibit a Fresnel zone structure with vanishing sensitivity along the ray path, known from finite-frequency traveltime measurements on seismometer recordings \citep[e.g.][]{Dahlen_2000}. In accord with Eq. \ref{E:000}, the kernels connect the source to segments of the fiber with large curvature.\\
Modifying the source from an explosion to a moment tensor with zero trace, generates S waves in addition to P waves. As illustrated in Fig. \ref{F:kernels_2}, they also appear in the phase change time series $\theta(t)$ as a sequence of oscillations produced by different fiber segments. The fiber segment responsible for a specific oscillation can again be determined by sensitivity kernel analysis, now providing sensitivity with respect to both P and S velocity. Fig. \ref{F:kernels_2} corroborates that segments with large curvature behave similar to discrete, i.e., distributed measurement points, thereby providing window-specific sensitivity coverage, despite the integrated nature of the phase change time series.
%
%
%
%
%
%

\section{Discussion and Conclusions}\label{S:Conclusions}

Using adjoint techniques for structural sensitivity analysis, we have shown with conceptual examples that optical sensing systems based on the transmission of deformation-induced phase changes can effectively be used to make space-distributed measurements. The extent to which this is possible depends on various factors. Most importantly, the fiber must contain curved segments that behave similar to localized sensors because the sensitivity of a fiber segment to deformation is proportional to its local curvature $\frac{d}{ds}\w{e}(s)$. Provided that the spacing of strongly curved segments is larger than a seismic wavelength, they can produce a sequence of distinguishable wavelets in the phase change time series $\theta(t)$, each representing local, i.e., distributed, deformation. Finally, according to Eq. \ref{E:000}, the fiber normal vector $\frac{d}{ds}\w{e}(s)$ should be roughly parallel to the displacement field polarization $\w{u}$.\\
In the presence of a suitably shaped fiber, the time-dependent analysis of $\theta(t)$ may provide a set of traveltime (or other) measurements and sensitivity kernels that are useful for the solution of tomographic inverse problems. In this context, strongly curved fiber segments roughly mimic a network of conventional seismic instruments. This may, indeed, be beneficial for imaging remote regions such as ocean basins, where telecommunication cables already exist.\\
It must, unfortunately, be suspected, that developers of future fiber-optic infrastructure might ignore the wishes of structural seismologists and deploy cables in accord with economic and political boundary conditions, instead of maximizing curvature. Nevertheless, the methods presented in section \ref{S:Theory} provide the theoretical foundation for optimal experimental cable design, as well as tools for the sensitivity analysis of existing fiber-optic cable installations.


\begin{acknowledgments}
Andreas Fichtner gratefully acknowledges United Airlines for a 7-hour delay at New Orleans airport, which provided ample time to develop the forward modelling theory, summarized in Eq. (\ref{E:000}).
\end{acknowledgments}


\bibliography{biblio}
 
\end{document}